\documentclass{article}

\usepackage[english]{babel}
\usepackage[utf8]{inputenc}
\usepackage{amsmath}
\usepackage{amsthm}
\usepackage{tikz}
\tikzset{>=latex}
\usetikzlibrary{decorations,arrows,shapes}
\usepackage[round]{natbib}
\usepackage{authblk}
\providecommand{\keywords}[1]{\textit{\textit{Keywords:}} #1}

\title{On weighted parametric tests}

\author{Dong Xi, Ekkehard Glimm, Willi Maurer, Frank Bretz}

\affil{Statistical Methodology, Novartis}

\date{\today}

\begin{document}
\maketitle

\begin{abstract}
We describe a general framework for weighted parametric multiple test procedures based on the closure principle. We utilize general weighting strategies that can reflect complex study objectives and include many procedures in the literature as special cases. The proposed weighted parametric tests bridge the gap between rejection rules using either adjusted significance levels or adjusted $p$-values. This connection is possible by allowing intersection hypotheses to be tested at level smaller than $\alpha$, which may be needed for certain study considerations. For such cases we introduce a subclass of exact $\alpha$-level parametric tests which satisfy the consonance property. When only subsets of test statistics are correlated, a new procedure is proposed to fully utilize the parametric assumptions within each subset. We illustrate the proposed weighted parametric tests using a clinical trial example.
\end{abstract}

\keywords{Multiple test procedure, Closure principle, Adjusted $p$-value, Non-exhaustiveness, Consonance
}

\section{Introduction}

Scientific experiments are often faced with simultaneous inference problems when addressing multiple objectives, such as assessing the differences between several experimental conditions. Weighted multiple test procedures (MTPs) are commonly used to control the overall Type \rm{I} error rate by assigning weights to different hypotheses in order to reflect the relative importance of objectives in the test strategy. For example, early references on weighted min-$p$ tests include the resampling-based tests from \citet[Chapter 6]{westfall1993resampling}, and \citet{westfall1998using}. Weighted MTPs based on specific parametric models have been investigated using hierarchical tests \citep{huque2008flexible} and graphical approaches \citep{bretz2011graphical}. These procedures discuss weighted parametric MTPs using the closure principle \citep{marcus1976closed} where each intersection hypothesis is tested at exact level $\alpha$. When there are no logical restrictions among the hypotheses, such as for the step-down Dunnett procedure \citep{dunnett1991step}, a weighted parametric test has been introduced by \citet{xie2012weighted} based on adjusted $p$-values.

MTPs are usually carried out by comparing either adjusted significance levels with unadjusted $p$-values or adjusted $p$-values with the unadjusted level $\alpha$. Although various weighted parametric tests have been proposed in the literature, the link between rejection rules using either adjusted significance levels or adjusted $p$-values has not been systematically explored. In addition, the majority of the procedures in the literature focus on the case where each intersection hypothesis is tested at exact level $\alpha$. It remains unclear how to deal with the non-trivial case where the significance level is strictly less than $\alpha$ for some of the intersection hypotheses. This is a relevant question for certain study considerations. For example, in the phase III clinical trial of buparlisib in patients with advanced and metastatic breast cancer, the analysis of progression-free survival (PFS) endpoints happens much earlier in time than the analysis of the overall survival (OS) endpoints. Thus, testing of PFS hypotheses does not benefit from rejecting the OS hypotheses at a later time point. \citep{goteti2014some}. Besides, in certain parallel and $k$-out-of-$n$ gatekeeping procedures, some intersection hypotheses involving primary hypotheses are tested at level smaller than $\alpha$ to allow testing secondary hypotheses if a certain number of primary hypotheses have been rejected \citep{dmitrienko2008general,xi2014general}.

We propose a unified framework for weighted parametric MTPs using the closure principle. This framework allows for general weighting strategies and includes many procedures in the literature as special cases. When some intersection hypotheses are tested at level smaller than $\alpha$, we reveal a special property of a class of parametric tests which proportionally increases the hypothesis weights to ensure exact $\alpha$-level tests. When the parametric assumptions only apply to subsets of hypotheses, we propose a new procedure which utilizes the parametric assumptions within each subset. We derive analytic expressions for the adjusted $p$-values to avoid numerical root finding under multidimensional integration.

\section{Notation}
\label{sec:notation}
Consider testing $m$ elementary null hypotheses $H_i,i\in I=\{1,\ldots,m\}$. Under the closure principle \citep{marcus1976closed}, we test each non-empty intersection hypothesis $H_J=\cap_{j\in J}H_j$, $J\subseteq I$, at level $\alpha$. We reject an elementary hypothesis $H_i$, $i\in I$, if every intersection hypothesis $H_J$ with $i\in J\subseteq I$ is rejected by its associated $\alpha$-level test. The closed procedure controls the familywise error rate (FWER) at level $\alpha$ in the strong sense \citep{hochberg1987multiple}.

Because some hypotheses among $H_1,\ldots,H_m$ may be more important than others, we assign weights for different hypotheses to reflect the relative importance. Using the notation from \citet{maurer2013memory}, let $w_{J} = (w_j(J), j \in J)$ denote a vector of weights for an index set $J \subseteq I$. A weighting scheme $W = \{w_{J}, J \subseteq I\}$ is called valid if for every $J\subseteq I$ and $j\in J$ we have $w_j(J)\geq 0$ and $0< \sum_{j \in J} w_j(J) \leq 1$. Validity is a basic but important condition and thus all weighting schemes considered in this paper are valid. In addition, $W$ is called exhaustive if for every $J\subseteq I$ we have $\sum_{j \in J} w_j(J) = 1$. Exhaustiveness is a desirable property but not required in this paper. For example, the weighting scheme of the step-down Dunnett procedure is $w_j(J)=1/\left\vert J \right\vert$ for $j\in J\subseteq I$, where $\left\vert J \right\vert$ denotes the number of indices in $J$.

Let $p_i$ denote the unadjusted $p$-value for $H_i$, $i\in I$. Consider the weighted Bonferroni test that rejects $H_J$ at level $\alpha$ if $p_j\leq w_j(J)\alpha$ for any $j\in J$. In the following, $w_j(J)$ and $w_j(J)\alpha$ are called the local weight and local significance level, respectively. An equivalent way of testing $H_J$ is to use its $p$-value $\hat{p}_J=\min[1,\min_{j\in J}\{p_j/w_j(J)\}]$. Accordingly, we can reject $H_J$ if $\hat{p}_J\leq \alpha$. Applying the closure principle, we can then reject the elementary hypothesis $H_i$ if its adjusted $p$-value $\max_{\{J:i\in J\subseteq I\}}\hat{p}_J \leq \alpha$.

Throughout this paper, we assume that under the null hypothesis $H_i$ the unadjusted $p$-value $p_i$ is uniformly distributed over $[0,1],i=1,\ldots,m$. The test statistic associated with $H_i$ is a function of $p_i$ under the inverse of the cumulative distribution function, which could be, for example, an (asymptotically) normal or a $t$ distribution. The joint distribution of the $p_i$'s is available if the corresponding test statistics follow a multivariate probability distribution, such as an (asymptotically) multivariate normal distribution.

\section{Weighted parametric tests for intersection hypotheses}
\subsection{Joint distribution fully known}
\label{sec:parametric}
Let $P_j$ denote the random variable whose realization is the observed unadjusted $p$-value $p_j$ for $H_j,j\in J$, for some $J\subseteq I$. If the joint distribution of $P_j$, $j \in J$, is fully known, the weighted min-$p$ test rejects $H_J$ if $p_j\leq c_J w_{j}(J)\alpha$ for any $j\in J$, where $c_J$ is calculated such that
\begin{equation}
\text{pr}_{H_J}\left[\bigcup_{j\in J}\left\{P_j\leq c_J w_{j}(J)\alpha\right\}\right]=\alpha\sum_{j\in J}w_{j}(J).
\label{eq:test}
\end{equation}
Setting $c_J=1$ results in the weighted Bonferroni test with an inequality in \eqref{eq:test}. Otherwise, $c_J>1$ and the resulting weighted parametric test is more powerful than the weighted Bonferroni test. Let $q_J=\min_{j\in J}\left\{p_j/w_j(J)\right\}$ denote the smallest observed weighted $p$-value for $H_j$, $j\in J$. The $p$-value $\hat{p}_J$ for the intersection hypothesis $H_J$ subject to $\sum_{j\in J}w_{j}(J)\leq 1$ is then given by
\begin{equation}
\hat{p}_{J}=\min\left[1,\frac{1}{\sum_{j\in J}w_{j}(J)}\text{pr}_{H_J}\left[\bigcup_{j\in J}\left\{\frac{P_j}{w_j(J)}\leq q_J\right\} \right]\right].
\label{eq:adjp}
\end{equation}
Therefore, we reject $H_J$ if $p_j\leq c_J w_{j}(J)\alpha$ for any $i\in J$ with $c_J$ determined in \eqref{eq:test} or, equivalently, if $\hat{p}_J\leq \alpha$. By the closure principle, we reject an elementary hypothesis $H_i,i\in I$, if every $H_J$ with $i\in J\subseteq I$ is rejected. Equivalently, the adjusted $p$-value of $H_i$ is the maximum of $\hat{p}_J,i\in J\subseteq I$, and we reject $H_i$ if it is less than or equal to $\alpha$. Together with a general weighting scheme $W$, the proposed weighted parametric test \eqref{eq:test} and \eqref{eq:adjp} includes many procedures in the literature as special cases, such as the step-down Dunnett procedure \citep{dunnett1991step}, the parametric fallback procedure \citep{huque2008flexible}, and the graphical approaches with parametric assumptions \citep{bretz2011graphical}.

To see how $\hat{p}_{J}$ is derived in \eqref{eq:adjp}, rewrite the left hand side of \eqref{eq:test} as
$$
\text{pr}_{H_J}\left[\bigcup_{j\in J}\left\{\frac{P_j}{w_j(J)}\leq c_J \alpha\right\}\right]=\text{pr}_{H_J}\left[\min_{j\in J}\left\{\frac{P_j}{w_j(J)}\right\}\leq c_J \alpha\right]=\alpha\sum_{j\in J}w_{j}(J).
$$
Then $c_J\alpha$ is the $\left\{\alpha\sum_{j\in J}w_{j}(J)\right\}$th quantile of the distribution of the minimum weighted $p$-value $Q_{J}=\min_{j\in J}\left\{P_j/w_j(J)\right\}$. Under the null hypothesis $H_J$, the probability of observing an equally or more extreme outcome is $\text{pr}_{H_J}\left\{Q_{J} \leq q_J \right\}$. The $p$-value for $H_J$ subject to $\sum_{j\in J}w_j(J)\leq 1$, is then given by \eqref{eq:adjp}, after truncation at 1. Note that it is computationally more efficient to derive rejection rules using $\hat{p}_J$ because it avoids solving numerically for $c_J$ from an equation involving multidimensional integration.

\subsection{Parametric tests that enforce exhaustiveness}
\label{sec:exhaustive}
In Section \ref{sec:parametric}, we investigated weighted parametric tests that preserve the significance level for $H_J,J\subseteq I$, at level $\alpha\sum_{j\in J}w_j(J)$. However, it may be tempting to always increase the sum of the local weights to 1. \citet{xie2012weighted} considered the case when the initial weights $w_i=w_i(I)>0$ for all $i\in I$. If the joint distribution among the $p$-values is fully known, they proposed a closed procedure using
\begin{equation}
\hat{p}_{J}=\text{pr}_{H_J}\left[\bigcup_{j\in J}\left\{\frac{P_j}{w_j}\leq q_J\right\}\right],
\label{eq:adjpProp}
\end{equation}
where $q_J=\min_{j\in J}\left\{p_j/w_j\right\}$. Here, \eqref{eq:adjpProp} is stated more generally because we do not assume the ordering in weighted $p$-values as in Section 2$\cdot$4 of \citet{xie2012weighted}. Compared to \eqref{eq:adjp}, the factor $1/\sum_{j\in J}w_j(J)$ is missing, which implies that $\sum_{j\in J}w_j(J)$ is always increased to 1.

Note that Xie (2012) did not provide rejection rules based on adjusted significance levels. From the relationship between \eqref{eq:test} and \eqref{eq:adjp}, we can derive an equivalent rejection rule that $H_J$ is rejected if $p_j\leq c_J w_{j}(J)\alpha$ for any $j\in J$, where $c_J$ is calculated such that
\begin{equation}
\text{pr}_{H_J}\left[\bigcup_{j\in J}\left\{P_j\leq c_J w_{j}(J)\alpha\right\}\right]=\alpha=\text{pr}_{H_J}\left[\bigcup_{j\in J}\left\{P_j\leq c_J\frac{w_{j}}{\sum_{j\in J}w_j} \alpha\right\}\right].
\label{eq:testProp}
\end{equation}
If $w_j(J)=w_j/\sum_{j\in J}w_j$, the leftmost and the rightmost expressions in \eqref{eq:testProp} are the same. We then reject $H_J$ if $p_j\leq c_J \alpha w_{j}/\sum_{j\in J}w_j$
for any $j\in J$. Thus, the procedure by \citet{xie2012weighted} actually tests $H_J$ in the following two steps. First, set $w_j(J)$ to $w_j/\sum_{j\in J}w_j$, i.e., increase $w_j$ proportionally such that $\sum_{j\in J}w_j(J)=1$. Second, reject $H_J$ if $p_j\leq c_J \alpha w_{j}/\sum_{j\in J}w_j$ for any $j\in J$ as in \eqref{eq:testProp} or equivalently if $\hat{p}_J\leq \alpha$ as in \eqref{eq:adjpProp}.

The resulting weighting scheme is always exhaustive when $w_i>0$ for all $i\in I$. However, it requires that all local weights $w_j(J)$ are completely determined by the initial local weights, i.e., $w_j(J)=w_j/\sum_{j\in J}w_j,j\in J\subseteq I$. It does not apply to general weighting schemes, especially when some initial local weights are 0. Nevertheless, the idea by \citet{xie2012weighted} can be generalized to any valid weighting scheme by dropping $\sum_{i\in J}w_j(J)$ and $1/\sum_{i\in J}w_j(J)$ from the right hand side of \eqref{eq:test} and \eqref{eq:adjp}, respectively. The resulting closed procedure then always increases the local weight $w_j(J)$ proportionally to $w_j(J)/\sum_{i\in J}w_j(J)$.

It is not trivial to determine whether a weighting scheme generated from an MTP is exhaustive or not, even if the initial local weights sum to 1. In addition, it may be desirable to use a non-exhaustive weighting scheme for practical considerations. For these reasons, we recommend working on the weighting scheme separately to incorporate trial design considerations, and then using a weighted parametric test that preserves the significance level for each intersection hypothesis as in \eqref{eq:test} and \eqref{eq:adjp}. For instance, when $\sum_{i\in I}w_i<1$, the procedure by \citet{xie2012weighted} can be implemented by first proportionally increasing local weights so that the weighting scheme is $W=\left\{ w_{J}=\left(w_j/\sum_{j\in J}w_j, j\in J\right), J\subseteq I\right\}$. Then we can apply the weighted parametric test in Section \ref{sec:parametric} within the closed procedure.

\subsection{Joint distribution not fully known}
\label{sec:partial}
If the joint distribution is only known for subsets of $p$-values, we can extend the parametric test in \eqref{eq:test} and \eqref{eq:adjp} using ideas from \citet{bretz2011graphical}. Assume that $I$ can be partitioned into $\ell$ mutually exclusive subsets $I_h$ such that $I=\cup_{h=1}^{\ell}I_h$. For each subset $I_h$, $h=1,\ldots,\ell$, we assume that the joint distribution of the $p$-values $p_i$, $i\in I_h$, is fully known, but the joint distribution of $p$-values from different subsets is not necessarily known. For any $J\subseteq I$, let $J_h=J\cap I_h, h=1,\ldots,\ell$. Then we reject $H_J$ if $p_j\leq c_{J}w_{j}(J)\alpha$ for any $j\in J$, where $c_{J}$ is calculated such that
\begin{equation}
\sum_{h=1}^{\ell}\text{pr}_{H_J}\left[\bigcup_{j\in J_h}\left\{P_j\leq c_{J} w_{j}(J)\alpha\right\}\right]=\alpha\sum_{i\in J}w_{j}(J).
\label{eq:testPartialCommon}
\end{equation}
The approach from \citet{bretz2011graphical} is a special case of \eqref{eq:testPartialCommon} when $\sum_{j\in J}w_j(J)=1$.

Note that \eqref{eq:testPartialCommon} uses a common $c_J$ for all subsets $J_h, h=1,\ldots,\ell$. Hence, the test decisions in $J_h$ are affected by the distribution in other subsets although the joint distribution between subsets is not necessarily known. For example, if $J_h=\{j\}$ contains only one index, we reject $H_J$ if $p_j\leq c_{J} w_{j}(J)\alpha$, which is no longer the rejection rule if the Bonferroni test were applied. Instead, we propose to use different $c_{J_h}$'s for different subsets $J_{h}, h=1,\ldots,\ell$, to fully utilize the parametric assumptions for $J_h$. Specifically, for any $J\subseteq I$, we reject $H_J$ if $p_j\leq c_{J_h}w_{j}(J)\alpha$ for any $j\in J$, where $c_{J_h}$ is calculated such that
\begin{equation}
\text{pr}_{H_J}\left[\bigcup_{j\in J_h}\left\{P_j\leq c_{J_h} w_{j}(J)\alpha\right\}\right]=\alpha\sum_{j\in J_h}w_{j}(J)
\label{eq:testPartial}
\end{equation}
for $h=1,\ldots,\ell$. If we take the sum of the left hand side in \eqref{eq:testPartial} over $h=1,\ldots,\ell$, we have $\alpha\sum_{j\in J}w_{j}(J)$ on the right hand side, which is the significance level for $H_J$.

Another advantage of using different $c_{J_h}$'s for $J_h$ is that we can derive the $p$-values analytically. First, the $p$-value for each subset $J_h$ is derived using \eqref{eq:adjp} and then the $p$-value for $H_J$ is the minimum over $h=1,\ldots,\ell$. Specifically, let $q_{J_h}=\min_{j\in J_h}\{p_j/w_j(J)\}$ such that the $p$-value for $H_J$ becomes $\hat{p}_{J}=\min_{h=1}^{\ell}\{\hat{p}_{J_h}\}$, where
\begin{equation}
\hat{p}_{J_h}=\min\left[1,\frac{1}{\sum_{j\in J_h}w_{j}(J)}\text{pr}_{H_J}\left[\bigcup_{j\in J_h}\left\{\frac{P_j}{w_j(J)}\leq q_{J_h}\right\}\right]\right].
\label{eq:adjpPartial}
\end{equation}

\section{Consonance}
\label{sec:consonance}
A closed procedure is called consonant \citep{gabriel1969simultaneous} if the rejection of $H_J,J\subseteq I$, further implies that at least one $H_j$, $j\in J$, is rejected. Consonance is a desirable property leading to a short-cut procedure which gives the same rejection decisions as the original closed procedure but with fewer operations to the order of $m$ or $m^2$ \citep{grechanovsky1999closed}. \citet{hommel2007powerful} proved that the monotonicity condition $w_j(J) \leq w_j(J')$ for all $j\in J' \subseteq J \subseteq I$, guarantees consonance if weighted Bonferroni tests are applied to all intersection hypotheses.

If a weighted parametric test is applied as in \eqref{eq:test}, \citet{bretz2011graphical} showed that
\begin{equation}
c_J w_j(J) \leq c_{J'} w_j(J') \text{ for all } j\in J' \subseteq J
\label{eq:monotonicP}
\end{equation}
ensures consonance. If \eqref{eq:monotonicP} is satisfied, Algorithm 3 in \citet{bretz2011graphical} carries out the short-cut procedure. But \eqref{eq:monotonicP} is not always satisfied even if $w_j(J) \leq w_j(J')$ for all $j\in J' \subseteq J\subseteq I$. In such cases, \citet{bretz2011graphical} proposed to modify the weighting scheme such that \eqref{eq:monotonicP} is satisfied for a particular significance level $\alpha$. However, to calculate the $p$-values \eqref{eq:adjp} for $H_J$, this modification has to be satisfied for $c_J$ under all $\alpha\in [0,1]$, which is difficult to achieve.

The procedure by \citet{xie2012weighted} considers a special weighting scheme that ensures consonance. As in Section \ref{sec:exhaustive}, it assumes $w_i>0$ for all $i\in I$ and defines the weighting scheme as $w_j(J)=w_j/\sum_{j\in J}w_j$, $j\in J\subseteq I$. If the joint distribution of all test statistics is fully known, we calculate $c_{J}$ and $c_{J'}$ such that
$$
\text{pr}_{H_J}\left[\bigcup_{j\in J}\left\{p_j\leq c_J \frac{w_j}{\sum_{j\in J}w_j}\alpha\right\}\right]=\alpha=\text{pr}_{H_{J'}}\left[\bigcup_{j\in J'}\left\{p_j\leq c_{J'} \frac{w_j}{\sum_{j\in J'}w_j}\alpha\right\}\right].
$$
Because $J'\subseteq J$, the above equalities can only hold if $c_J /\sum_{j\in J}w_j\leq c_{J'}/\sum_{j\in J'} w_j$, which leads to \eqref{eq:monotonicP}. \citet{xie2012weighted} also made a similar assessment using $p$-values but did not refer to consonance explicitly. In fact, \eqref{eq:monotonicP} continues to hold even when the joint distribution of all test statistics is not fully known, as in \eqref{eq:testPartialCommon}.

\citet{xie2012weighted} provided a short-cut procedure to calculate the adjusted $p$-value for each elementary hypothesis. Here, we simplify the algorithm and do not assume the ordering in the weighted unadjusted $p$-values. For the overall intersection hypothesis $H_{J_1},J_1=I$, we calculate its $p$-value $\hat{p}_{J_1}$ according to \eqref{eq:adjpProp}. If $\hat{p}_{J_1}\leq \alpha$, reject $H_{j_1}$ with the adjusted $p$-value $\hat{p}_{j_1}=\hat{p}_{J_1}$ and proceed to the next step, where $j_{1}=\text{argmin}_{j\in J_1}\ p_j/w_j$; otherwise stop. In general, for $i=2,\ldots,m$, let $J_i=J_{i-1}\setminus \{j_{i-1}\}$ and calculate the $p$-value $p_{J_i}$ for $H_{J_i}$. If $p_{J_i}\leq \alpha$, reject $H_{j_i}$ with the adjusted $p$-value $\hat{p}_{j_i}=\max\{\hat{p}_{i_{i-1}},\hat{p}_{J_i}\}$, and proceed to the next step (as long as $i<m$), where $j_{i}=\text{argmin}_{j\in J_i}\{p_j/w_j\}$; otherwise stop. This short-cut procedure is performed in at most $m$ operations and can be viewed as a weighted version of the step-down Dunnett procedure.

If we generalize the procedure by \citet{xie2012weighted} to any valid weighting scheme, a sufficient condition for \eqref{eq:monotonicP} is that $w_j(J)$ can be written as $c_J w_j$, where $c_J$ is a constant for all $j\in J$. As a simple example, we derive a weighted version of the single-step \citet{dunnett1955multiple} test. The weighting scheme for the closed procedure is $W=\{w_{J}=(c_I w_j,j\in J),J\subseteq I\}$ such that $H_J$ is rejected if $p_j\leq c_I w_j \alpha$ for any $j\in J\subseteq I$. Here, $c_I$ is a constant for every $j\in J\subseteq I$ such that $\text{pr}_{H_I}\left\{\cup_{j\in I}\left(P_j\leq c_I w_{j}\alpha\right)\right\}=\alpha$. Assuming $\sum_{i\in I}w_i=1$, we derive the short-cut procedure for the weighted single-step Dunnett test which rejects $H_i$ if $p_i \leq c_I w_i\alpha$ for any $i\in I$. The adjusted $p$-value for $H_i$ is $\hat{p}_{i}=\text{pr}_{H_I}\left\{\cup_{j\in I}\left(P_j/w_j \leq p_i/w_i\right) \right\}$. Then the single-step \citet{dunnett1955multiple} test is a special case when $w_i=1/m,i\in I$.

\section{Clinical trial example}
\label{sec:example}
\begin{figure}[htbp]
\centering
\begin{tikzpicture}[scale=0.6]
\node at (70bp,-125bp) {Efficacy};
\node at (125bp,-80bp) {0$\cdot$4};
\node (HE1) at (125bp,-125bp)[draw,circle,inner sep=2pt,minimum size=2pt] {$H_1$};
\node at (325bp,-80bp) {0$\cdot$4};
\node (HE2) at (325bp,-125bp)[draw,circle,inner sep=2pt,minimum size=2pt] {$H_2$};
\node at (525bp,-80bp) {0$\cdot$2};
\node (HE3) at (525bp,-125bp)[draw,circle,inner sep=2pt,minimum size=2pt] {$H_3$};
\node at (70bp,-225bp) {Safety};
\node at (125bp,-270bp) {High dose};
\node at (160bp,-230bp) {0};
\node (HS1) at (125bp,-225bp)[draw,circle,inner sep=2pt,minimum size=2pt] {$H_4$};
\node at (325bp,-270bp) {Medium dose};
\node at (360bp,-230bp) {0};
\node (HS2) at (325bp,-225bp)[draw,circle,inner sep=2pt,minimum size=2pt] {$H_5$};
\node at (525bp,-270bp) {Low dose};
\node at (560bp,-230bp) {0};
\node (HS3) at (525bp,-225bp)[draw,circle,inner sep=2pt,minimum size=2pt] {$H_6$};
\draw [->,line width=0.5pt] (HE1) to node[pos=0.5,left] {1} (HS1);
\draw [->,line width=0.5pt] (HE2) to node[pos=0.5,left] {1} (HS2);
\draw [->,line width=0.5pt] (HE3) to node[pos=0.5,right] {1} (HS3);
\draw [->,line width=0.5pt] (HS1) to node[pos=0.3,below] {0$\cdot$5} (HE2);
\draw [->,line width=0.5pt] (HS1.50) arc(140:117:317bp) arc(117:70:317bp) to (HE3);
\node at (225bp,-125bp) {0$\cdot$5};
\draw [->,line width=0.5pt] (HS2) to node[pos=0.3,below] {0$\cdot$5} (HE1);
\draw [->,line width=0.5pt] (HS2) to node[pos=0.3,below] {0$\cdot$5} (HE3);
\draw [->,line width=0.5pt] (HS3.130) arc(40:63:317bp) arc(63:110:317bp) to (HE1);
\node at (425bp,-125bp) {0$\cdot$5};
\draw [->,line width=0.5pt] (HS3) to node[pos=0.3,below] {0$\cdot$5} (HE2);\end{tikzpicture}
\caption{Graphical multiple test procedure for the clinical trial example.}
\label{fig:example}
\end{figure}
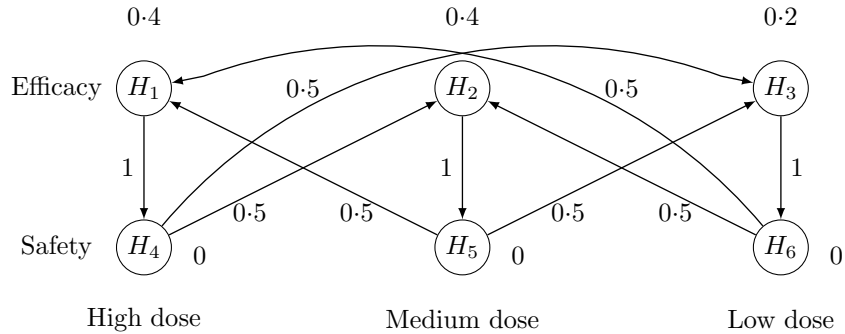

Consider the clinical trial example from \citet{bauer2001multiple} to test for the superiority of three doses of an investigational treatment against a control regarding an efficacy and a safety endpoint. There are three efficacy hypotheses $H_1,H_2,H_3$ and three safety hypotheses $H_4,H_5,H_6$ for the comparison of the high, medium, low dose against the control, respectively. We modify the step-down procedure without order constraints between the doses from Section 3 in \citet{bauer2001multiple} as follows. Assume the initial weights as $w_I=(\text{0$\cdot$4},\text{0$\cdot$4},\text{0$\cdot$2},0,0,0)$. Within each dose-control comparison, the hypothesis on the efficacy endpoint is tested first and, if rejected, the test on the safety endpoint is performed at the same local significance level. If both hypotheses can be rejected for a same dose, the associated local level is equally distributed among the other two doses. The one-sided significance level is $\alpha=\text{0$\cdot$025}$. The graphical representation of this MTP is shown in Figure \ref{fig:example} using the graphical approach from \citep{bretz2009graphical,burman2009recycling}. In this framework, hypotheses are denoted by nodes associated with their local weights. A directed edge from $H_i$ to $H_j$ means that when $H_i$ is rejected, its local weight can be propagated to $H_j$. The number associated with the edge quantifies the proportion of the local weight of $H_i$ that can be propagated to $H_j$.

The weighting scheme of this MTP can be obtained using Algorithm 1 in \citet{bretz2011graphical}, which has been implemented in the {\tt gMCP} R package \citep{Rohmeyer2015}. Table 1 in the Supplementary Material provides the local weight vector $w_J$ for each intersection hypothesis $H_J, J \subseteq I$. For example, the local weights for $H_{123}=H_1\cap H_2\cap H_3$ are $w_1=\text{0$\cdot$4},w_2=\text{0$\cdot$4}$ and $w_3=\text{0$\cdot$2}$, and for $H_{234}=H_2\cap H_3\cap H_4$ they are $w_2=\text{0$\cdot$4},w_3=\text{0$\cdot$2}$ and $w_4=\text{0$\cdot$4}$. The \citet{dunnett1955multiple} test is suitable for the many-to-one comparisons as it is more powerful than the Bonferroni test. This motivates us to use a weighted parametric test for the intersection hypothesis involving any two of the three efficacy hypotheses $H_1,H_2,H_3$. For the sake of illustration, we assume that the joint distribution of test statistics between the safety hypotheses $H_4,H_5,H_6$ is unknown.

We assume that the joint distribution of the test statistics for $H_1,H_2,H_3$ is trivariate normal with a mean vector of 0's. Assuming equal group sizes, the pairwise correlation between the test statistics is 0$\cdot$5 among the efficacy hypotheses. All other correlations are assumed to be unknown. Given this joint distribution, the index set $I=\{1,\ldots,6\}$ is partitioned into four subsets: $\{1,2,3\}$, $\{4\}$, $\{5\}$, $\{6\}$. Within the subset $\{1,2,3\}$, the test statistics follow the trivariate normal distribution.

We calculate the local significance levels for all intersection hypotheses using (A) the weighted Bonferroni test, (B) the weighted parametric test \eqref{eq:testPartialCommon} and (C) the weighted parametric test \eqref{eq:testPartial}. Using (A), the local significance levels for $H_{234}$ are $(\text{0$\cdot$4},\text{0$\cdot$2},\text{0$\cdot$4})\times \text{0$\cdot$025}=(\text{0$\cdot$01},\text{0$\cdot$005},\text{0$\cdot$01})$. Using (B), we calculate $c_{234}=\text{1$\cdot$033}$ from \eqref{eq:testPartialCommon} via the {\tt mvtnorm} package in R \citep{Genz2016}. The resulting local significance levels are $(\text{0$\cdot$4},\text{0$\cdot$2},\text{0$\cdot$4})\times \text{1$\cdot$033}\times \text{0$\cdot$025}=(\text{0$\cdot$0103},\text{0$\cdot$0052},\text{0$\cdot$0103})$. Using (C), we calculate $c_{23}=\text{1$\cdot$057}$ and $c_{4}=1$ from \eqref{eq:testPartial}. The resulting local significance levels are $(\text{0$\cdot$4}\times \text{1$\cdot$057},\text{0$\cdot$2}\times \text{1$\cdot$057},\text{0$\cdot$4}\times 1) \times \text{0$\cdot$025}=(\text{0$\cdot$0106},\text{0$\cdot$0053},\text{0$\cdot$01})$.

From the above calculations we can see that both parametric tests (B) and (C) produce higher local significance levels than the Bonferroni test (A). Thus, they can reject at least as many hypotheses as the Bonferroni test. Differences between (B) and (C) arise when two efficacy hypotheses and at least one safety hypothesis with a positive weight are associated with an intersection. For example, for $H_{234}$ (C) preserves the level for the safety hypotheses at the level of the Bonferroni test (A) but produces higher level for the efficacy hypotheses than (B). On the other hand, (B) has a higher level for the safety hypotheses but a lower level for the efficacy hypotheses. These conclusions apply to all intersection hypotheses.

\section*{Supplementary material}
Supplementary material includes the weighting scheme and further numerical comparisons for the clinical trial example in Section \ref{sec:example}.

\bibliographystyle{unsrtnat}
\bibliography{paper-ref}

\end{document}